\documentclass[a4paper,12pt]{article}
\usepackage{amsmath}
\usepackage{amsfonts}
\usepackage{amssymb}
\usepackage[dvips]{graphicx}
\usepackage[T1]{fontenc}
\usepackage[utf8]{inputenc}
\usepackage{authblk}
\usepackage{epsfig}

\begin{document}

%\begin{frontmatter}
\title{Quasi-Topological Quantum Field Theories and $\mathbb{Z}_2$ Lattice Gauge Theories}

\author{\small{M. J. B. Ferreira}\thanks{migueljbf@fma.if.usp.br}}
\author{\small{V. A. Pereira}\thanks{vpalves@fma.if.usp.br}}
\author{\small{P. Teotonio-Sobrinho}\thanks{teotonio@fma.if.usp.br}}
\affil{\footnotesize{{\it Instituto de F\'isica, Universidade de S\~ao Paulo, \\ C.P. 05315-970, S\~ao Paulo-SP, Brazil.}}}

\date{\vspace{-.5cm}\small{May 28, 2012}}

\maketitle
\vspace{-1cm}

\abstract{
We consider a two parameter family of $\mathbb{Z}_2$ gauge theories on a lattice discretization $T({\cal M})$ of a 3-manifold ${\cal M}$  
and its relation to topological field theories. Familiar models such as the spin-gauge model 
are curves on a parameter space $\Gamma$. We show that there is a region $\Gamma_0 \subset \Gamma$
where the partition function and the expectation value $\langle W_R(\gamma)\rangle$ of the Wilson loop 
can be exactly computed. Depending on the point of $\Gamma_0$, the
model behaves as topological or quasi-topological. The partition function is, up to a scaling factor,
a topological number of ${\cal M}$. The Wilson loop on the other hand, does not depend on the 
topology of $\gamma$. However, for a subset of $\Gamma_0$, 
$\langle W_R(\gamma)\rangle$ depends on the size of $\gamma$ and follows a discrete version of an
area law. At the zero temperature limit, the spin-gauge model approaches the topological and the quasi-topological regions depending on the sign of the coupling constant. 
}

%\end{frontmatter}

\section{Introduction}\label{sec:Intro}

A lattice gauge theory with gauge group $\mathbb{Z}_2$ is the simplest example
of a gauge theory \cite{kogut}. In dimension $d=2$, the partition function
for $\mathbb{Z}_2$ (as for any other compact gauge group)
can be computed in various ways. In dimensions larger than two, however, the
simplicity of $\mathbb{Z}_2$ does not help us to solve the model. Even without matter, the relevant
models are non trivial and exact solutions are not known. Such solutions would be
a very important achievement.
A $\mathbb{Z}_2$ gauge theory on a cubic 
lattice can be made dual to the $3D$ Ising model, an outstanding problem in 
statistical mechanics \cite{wegner,savit}. It is clear that in $d=4$ the problem is at least as difficult
as in $d=3$. In this paper we will not have much to say about $d=4$ since the tools 
we use are peculiar to dimension three.

However difficult, $3D$ lattice models with local gauge symmetry are not always beyond the
reach of exact solutions. That depends on the dynamics, i.e., the choice of an action 
for a lattice plaquette. Topological Quantum Field Theories (TQFTs) are examples where one can
perform exact computations. Examples have been constructed on the
lattice in dimensions $d=3$ \cite{kuperberg,cfs} as well as $d=4$ \cite{4dtqft}. 
Once the topology on the manifold in question is fixed, the partition function does not
depend on the lattice size and can be trivially computed for a discretization with very small
number of sites, links and plaquettes. Such models are very simple from the physical point of view. 
It follows from topological invariance that transfer matrices are trivial. 

Despite being trivial dynamically for a fixed topology, TQFTs are
quite relevant in physics. The reason being that TQFTs can come out 
as limits of ordinary field theories in the continuum as well as in the lattice.
The most celebrated example is topological order in condensed matter physics \cite{topologicalorder}
where the physics at large scales is described by a TQFT. Something of similar nature
also happens for lattice theories in $d=2$ that are quasi-topological \cite{teot1,teot2}.
They are reduced to a TQFT at the appropriate limits. 

In order to understand the relationship between fully dynamical $d=2$ models and their
possible topological limits one can first look at quasi-topological models. 
They are very easy to work with since we can compute
all relevant quantities. Quasi-topological 
models in $d=2$ are a nice set of toy models for this purpose. 
In particular, the relation between the original models and their topological limits
is made very explicit. As for $d=3$ the situation is more
complicated. We do no have at our disposal toy models that are at the same time 
not topological and easily computable. We have no choice but
to work with fully dynamical theories where no exact 
computations are available.

The focus of this paper is to investigate how lattice field theories with local $\mathbb{Z}_2$ 
gauge symmetry are related to topological theories. Before going any further we need 
to say what we mean by a lattice model being topological. Let $T(\mathcal{M})$ be a 
lattice triangulation
of a fixed compact $3$-manifold $\mathcal{M}$. 
We say that such a model is topological if
the partition function is the same for all triangulations $T(\mathcal{M})$. Actually, this is a weak 
definition since we may ask that not only the partition function but the expectation
value of all observables to be of a topological nature. In any case, we have to go
beyond the usual regular cubic lattices and take into account arbitrary lattices.

In \cite{n1} we investigated the spin-gauge
model in $d=3$. 
%The theory is defined as follows. Let $\mathcal{M}$ be a compact $3$-manifold 
%and $T(\mathcal{M})$ a lattice discretization given by a triangulation of $\mathcal{M}$. 
%The gauge variables sitting at a link $a$ of a lattice $L$ are given by
%$g_a\in \mathbb{Z}_2=\{1,-1\}$. 
%Let $\omega(f)$ be the 
%oriented product of all gauge variables at the boundary of a particular face $f\in L(M)$. 
%Then the action is $S = \beta \sum_f \omega(f)$, where $\beta$ is the coupling constant 
%of the theory, and the sum runs over all faces $f$. The partition
%function is given by $Z=\sum_{\{\sigma_a\}} \exp(S)$, the sum running over all 
%gauge configurations $\{g_a\}$.  
We showed that in the limit $\beta\rightarrow \infty$ the partition function $Z$ is given by 
\begin{equation}
Z\left(T(\mathcal{M})\right)|_{\beta \rightarrow \infty} =
Z^{top}(\mathcal{M})~2^{N_F+N_L-N_T}, 
\label{eq:lowsol}
\end{equation}
where $N_T,N_F$ and $N_L$ are the number of tetrahedra, faces and links of $T(\mathcal{M})$ and
$Z^{top}(\mathcal{M})$ is a topological number and as such does not depend on the discretization $T(\mathcal{M})$. 
Equation (\ref{eq:lowsol}) tell us that at the limit $\beta \rightarrow \infty$ the 
partition function is not strictly speaking topological since it depends on the triangulation. However it does not
depend on the details of the triangulation but only on its size. For this reason we say that
the the partition function is quasi-topological.
It follows from (\ref{eq:lowsol}) that
the partition function can be computed for all triangulations. Let $T_0(\mathcal{M})$ be
a lattice triangulation where the numbers $N_{T_0}$, $N_{F_0}$ and $N_{L_0}$ are very small such that $Z(T_0)$
can be written down explicitly. For an arbitrary lattice $T(\mathcal{M})$ we have
\begin{equation}
Z(T)|_{\beta \rightarrow \infty}= Z(T_0)|_{\beta \rightarrow \infty}~2^{(N_F-N_{F_0})+(N_L-N_{L_0})-(N_T-N_{T_0})}.
\label{eq:beta=infty}
\end{equation}

We will find it convenient to rewrite 
$Z$ as a product of local Boltzmann weights, in the form
\begin{equation}
Z=\sum_{\{g_a\}} \prod_f W(f), \label{eq:Z2}
\end{equation}
where the produt is over all faces of the triangulation and sum is over all configurations. Let $a,b$ and $c$ be the links of a face $f$ and $(g_a,g_b,g_c)$ a gauge configuration at $f$. The corresponding local Boltzmann weight for the spin-gauge model can be written as
\begin{equation}
W^{(1)}(g_a,g_b,g_c)=e^{\beta g_ag_bg_c}.\label{eq:spin}
\end{equation}
Another very common choice is to set the local Boltzmann weight to
\begin{equation}
W^{(2)}(g_a,g_b,g_c)=e^{-\beta( 1-g_ag_bg_c)},\label{eq:wilson}
\end{equation}
which corresponds to the usual gauge theories where flat holonomies will
have the highest weight.

As we will see in this paper, the relationship between gauge models
and TQFTs can be better understood if we depart from
a specific example as in \cite{n1} and consider a more
general class of gauge models. In order to have a gauge theory, the local weight 
should depend only on the product of the gauge variables around an oriented plaquette: 
\begin{equation}
W(g_a,g_b,g_c)=M(g_ag_bg_c).
\end{equation}
Gauge invariance means that $M(g)$ is a class function or, in other words
$M(hgh^{-1})=M(g), \;\; \forall h\in G$. The character expansion for the group $\mathbb{Z}_2$ 
is very simple and implies that 
\begin{equation}
M(g_a,g_b,g_c)=m_1(g_ag_bg_c)+m_0 ~, \label{eq:theModel}
\end{equation}
where $m_i\in \mathbb{R}$. The original 
spin-gauge model with one parameter $\beta$ can be recovered by
restricting the model to a curve 
$( m_0,m_1 )=\left(\cosh\left( \beta \right),\sinh\left( \beta \right)\right)$ in this two-dimensional parameter space.
We will refer to the parameter space as $\Gamma$.

The first question to be addressed is the generalization of 
equation (\ref{eq:beta=infty}). We will show that there is a
subset $\Gamma_0$ of the parameter space $\Gamma$ such that the  partition function
$Z(m_0,m_1,T(\mathcal{M}))$ can be written as a product of a topological
invariant $Z^{top}(\mathcal{M})$ times a known function depending on the numbers $N_T$, $N_F$ and $N_L$ of tetrahedra, faces and links. Again, an easy consequence is that at $\Gamma_0$ 
the partition function $Z(m_0,m_1,T(\mathcal{M}))$ can be computed for any lattice $T(\mathcal{M})$. 
The subset $\Gamma_0$ is made of two pairs of lines, namely,
$m_i=0, i=0,1$ and  $m_1=\pm m_0$. 
It turns out that these two regions of $\Gamma_0$
have different properties. For instance, the topological invariant
that appears in the first pair is trivial and $Z^{top}(\mathcal{M})=1$
for all compact manifolds $\mathcal{M}$. As for second pair, $Z^{top}(\mathcal{M})$ depends
on the first group of co-homology of $\mathcal{M}$ \cite{n1}. The two pairs of solutions
are also related to high and low temperature limits as it will be clear
from the discussion on section \ref{sec:definitions}.

As in any gauge theory, one would be interested in more observables than just the
partition function. In particular it  is important to
calculate the expectation value $\left<W_R(\gamma)\right>$ of
Wilson loops for arbitrary representations $R$ of the gauge group and closed 
curves $\gamma$. In our previous work \cite{n1} we considered only the
partition function. In the present paper, we would 
like to go further and ask whether $\left<W_R(\gamma)\right>$ can be computed
for some points of $\Gamma$. As it happens for the partition function, 
such computation can be performed for all points of $\Gamma_0$. 
Note that in a truly topological
gauge theory such as Chern-Simons,  $\left<W_R(\gamma)\right>$ is a topological invariant
of $\gamma$. 
That is not true for all points of $\Gamma_0$.
It turns out that $\left<W_R(\gamma)\right>$ depends on the size of $\gamma$ 
for points of $\Gamma_0$ of the form $(m_0,m_1)=(\lambda,-\lambda)$.
Since there is no dependence on the
details of $\gamma$ we say that the observables
$\left<W_R(\gamma)\right>$ are quasi-topological.

Our approach is based on the fact that a large class of lattice
models, topological or otherwise, can be described by a set of
algebraic data on a  vector space $V$. These data comprises of a multiplication
$m$, a co-multiplication $\Delta$ and an endomorphism $S$ such that $S^2=1$.
It is also assumed that there is a unity $e$ and a co-unity $\epsilon$. 
It has been shown in \cite{kuperberg,cfs} that  when the data 
$(m,\Delta,S,e,\epsilon)$ defines a Hopf algebra, one can construct a
lattice topological field theory. We observed in \cite{n1} that 
the same data can be used to describe an ordinary $\mathbb{Z}_2$ gauge
theory. In \cite{n1}, however, $(m,\Delta,S,e,\epsilon)$ is not a 
Hopf algebra. In particular, the co-multiplication is not
an algebra morphism as it happens for Hopf algebras. Another important
difference is that instead of a fixed algebraic data, we had a
one parameter family $m(\beta)$ of multiplications where $\beta$ is
the coupling constant of  the model. It turns out that a Hopf
algebra is recovered in the limit $\beta\rightarrow \infty$ and
the model becomes quasi-topological. It is also possible to have examples with 
matter fields via a family of co-multiplications 
$\Delta(\lambda)$ where $\lambda$ is the corresponding coupling constant
\cite{n2}. In this paper, however, we will be limited to pure
gauge theories. As stated before, we will consider gauge theories 
with two coupling constants given by (\ref{eq:theModel}).
If we were to follow the formalism of \cite{n1}, that would
be encoded in a two
parameter family of multiplications 
$m(m_0,m_1)$. The model 
considered in  \cite{kuperberg,cfs} corresponds to the unique
point $(m_0,m_1)=(1,1)$ where  
$m(1,1)$  together with $\Delta,S,e,$ and $\epsilon$ define 
a Hopf algebra. For the
present paper, however, the Hopf structure is less
important. What matter are the points where the model
is quasi-topological. That happens for $(m_0,m_1)$ 
belonging to the region $\Gamma_0$ described above. 
The data $m(m_0,m_1),\Delta,S,e,$ defines a Hopf algebra
only at a single point of $\Gamma_0$.
 Just as
the topological model of \cite{kuperberg,cfs}, the limit $\beta \rightarrow \infty$
of equation (\ref{eq:lowsol}) also corresponds to a point in the
set $\Gamma_0$. 

The organization of the paper goes as follows: On Section
\ref{sec:definitions} we explain how the algebraic data can
be used to encode the model and how familiar models, such as the
spin-gauge model, fit into the parameter space $\Gamma$. 
On Section \ref{sec:limits} we 
determine the subset $\Gamma_0$ where
the model is quasi-topological. The computation
of the expectation value for Wilson loops is investigated on Section
\ref{sec:wilsonLoops}. We show that $\langle W_R(\gamma)\rangle $
can be computed for $\Gamma_0$. The model is not topological for all $\Gamma_0$.
We show that for a particular  region of $\Gamma_0$, $\langle W_R(\gamma)\rangle$
depends on the size of $\gamma$ and follows a discrete version of an area law.

We close the paper with some final remarks
on Section \ref{sec:final}.

\section{The Partition Function}\label{sec:definitions}

In this section we introduce the formalism we will use to 
describe the partition function of a 
$\mathbb{Z}_2$ gauge theories. 
We will loosely follow
\cite{kuperberg, cfs,n1} making the necessary modifications to suit
our purpose.

\begin{figure}[h!]
	\begin{center}
		\includegraphics[scale=1]{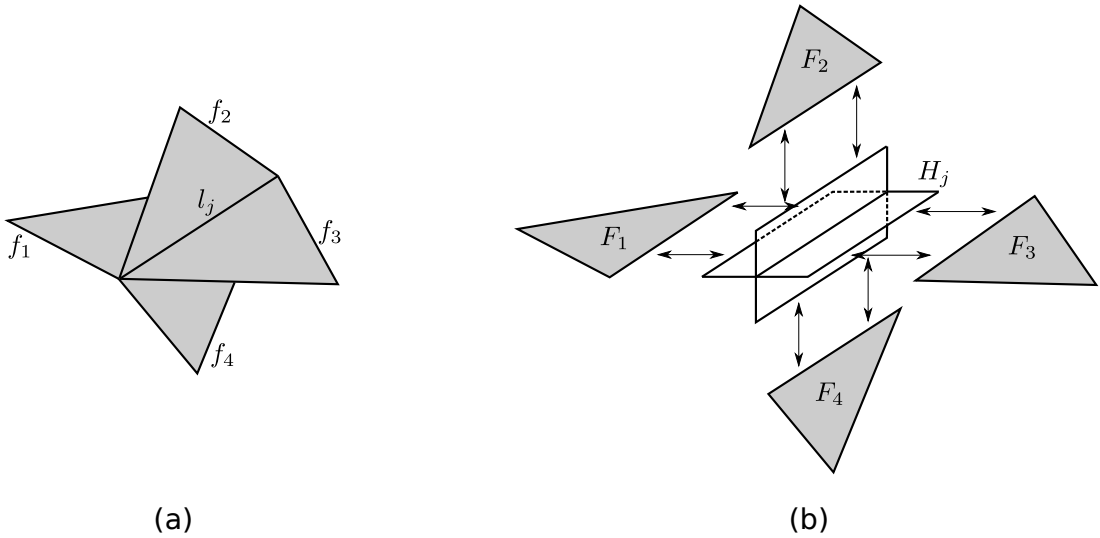}
		\caption{In {\bf (a)} we have a small piece of a triangulation. Figure {\bf (b)} shows its corresponding decomposition into flaps and faces.}
		\label{fig1}
	\end{center}
\end{figure}

Let $T(\mathcal{M})$ be a triangulation of a 
compact 3-dimensional  manifold $\mathcal{M}$.
For the description of a pure gauge theory, what is relevant in a discretization
is the set of faces and how they are interconnected. To encode the
connectivity information we will split $T(\mathcal{M})$ into individual faces
and record the information on how they should be put back together. 
This process can be described as follows. For each face $f_k\in T(\mathcal{M})$ 
we associate a disjoint
face $F_k$ and for each link $l_j\in T(\mathcal{M})$ we associate a hinge object
$H_j$ with $n_j$ flaps as illustrated in figure \ref{fig1}. The number
$n_j$ of flaps is equal to the number $I(l_j)$ of faces of 
$T(\mathcal{M})$ that share the link $l_j$. 
To reconstruct $T(\mathcal{M})$ from the disjoint
faces $F_k$, we can use the hinges $H_j$
to determine
which faces are to be joint together. This is illustrated by figure \ref{fig1}.
The faces $F_k$ and the hinges $H_j$ are to be given an extra structure
called orientation. For the case of $\mathbb{Z}_2$ gauge
theories, this orientations are not relevant but we will mention them as they help to organize the model.
We will call the set $\{F_k,H_j\}$ and its interconnections a decomposition
of $T(\mathcal{M})$.

\begin{figure}[h!]
	\begin{center}
		\includegraphics[scale=1]{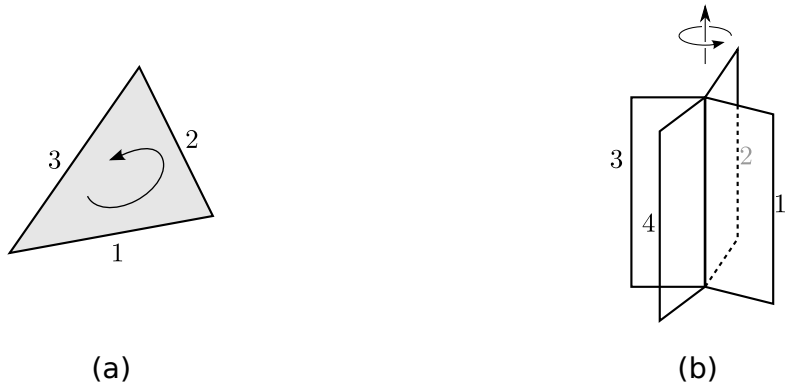}
		\caption{The links on a triangular face are numbered from 1 to 3. The orientation is indicated in {\bf (a)}. Figure {\bf (b)} shows a hinge with four flaps and its orientation.}
		\label{fig2}
	\end{center}
\end{figure}

We will use the decomposition $\{F_k,H_m\}$, plus some extra data, 
to define a partition function. The first step is to choose a
vector space $V$ of dimension $n$. The edges of a face $F_k$ have
to be enumerated from $1$ to $3$. That amounts to a choice of
orientation of the face and a choice of the starting point (see figure \ref{fig2} {\bf (a)}).
The edges of $l_m\in F_k$ carry configurations $(a^{k}_1,a_2^{k},a_3^{k})$,
with $a_i^{k}\in \{1,\cdots ,n\}$. A statistical weight 
$M_{a_1^{k}a_2^{k}a_3^{k}}$
will be associated to the face $F_k$.
Note that $M_{a^k_1b^k_2c^k_3}$ can be viewed as the components of a 
tensor $M\in V\otimes V\otimes V$. Furthermore, it
should be invariant by cyclic permutations
of its indices since we do not care which edge is to be numbered as the
first one. On the other hand, a change in orientation
can affect the corresponding weight since $M_{abc}$ may
not be the same as $M_{cba}$.
In a similar fashion, the flaps of a hinge $H_m$ can be cyclically
numbered from $1$ to $q=I(l_m)$. Once more, this is equivalent 
to give $H_m$ an orientation, as illustrated by figure \ref{fig2} {\bf (b)}.
The flaps of $H_m$
carry configurations 
$(a^m_1,\cdots ,a^m_{q})$, $a^m_i\in \{1,\cdots ,n\}$ just like the edges 
of a face. That will correspond to a statistical
weight $\Delta ^{a^m_1\cdots a^m_q}$. 

Such numbers can be interpreted as
the components of a tensor $\Delta \in V^*\otimes \cdots \otimes V^*$
and have to be invariant by cyclic permutations of the indices. As before,
change in orientation of $H_m$ will change the statistical weight to 
$\Delta ^{a^m_qa^m_{q-1}\cdots a^m_1}$.

\begin{figure}[h!]
	\begin{center}
		\includegraphics[scale=.8]{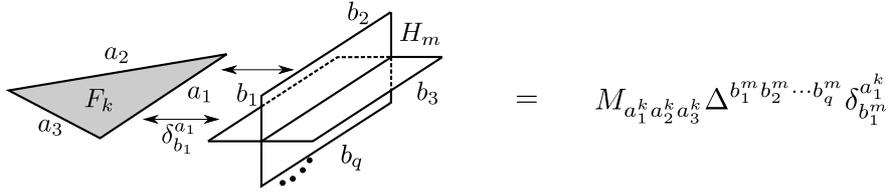}
		\caption{In terms of tensor, we interpret the gluings as contractions of the corresponding indices.}
		\label{fig3}
	\end{center}
\end{figure}

Once we fix an orientation for each $F_k$ and an orientation for each $H_m$, 
we produce a tensor $M_{a_1a_2a_3}(F_k)$ for each face and a 
tensor $\Delta^{b_1\cdots b_q}(H_m),q=I(l_m)$ for each hinge. One can see that
the product
\begin{equation}
\prod_k  M_{a^k_1a^k_2a^k_3}(F_k) \prod_m \Delta^{b^m_1\cdots b^m_q}(H_m),
\label{eq:big-tensor}
\end{equation}
has one covariant index for each edge and one contra-variant index for each flap
of $\{F_k,H_m\}$. The partition function will be the scalar constructed by 
contracting all indices. A covariant index $a^k_i$ is to be contracted to a contra-variant index
$b^m_j$ whenever the corresponding edge $F_k$ and flap $H_m$ are to be glued together.
In other words, we define the scalar $Z$ as
\begin{equation}
Z=\prod_k  M_{a^k_1a^k_2a^k_3}(F_k) \prod_m \Delta^{b^m_1\cdots b^m_q}(H_m) 
   \prod_{b\in \{gluings\}} \delta^{a^r_i}_{b^s_j}(b),
\label{eq:big-model}
\end{equation}
where the last product is responsible for contracting indices. There will be a 
$\delta^{a^r_i}_{b^s_j}(b)$ for each paring edge-flap $(F_r,H_s)$ that are glued together,
as illustrated by figure \ref{fig3}. We can simplify the notation of 
(\ref{eq:big-model}) by eliminating the Kronecker deltas and writing
\begin{equation}
Z=\prod_{f\in \{F\}}  M_{abc}(f) 
\prod_{l\in \{L\}} \Delta^{b_1\cdots b_{N_L}}(l), 
\label{eq:big-model-simple}
\end{equation}
where $\{F\}$ and $\{L\}$ are the set of faces and links of the triangulation.
A contraction on the indices corresponding to gluings is understood.

As of now, the partition function (\ref{eq:big-model}) is not very useful.
We need to be more precise about the weights $M_{a_1 a_2 a_3}(F_k)$ and $\Delta^{b_1\cdots b_n}(H_m)$
if we want $Z$ to be related to physical models like the spin-gauge model. 
Note that $Z$ depends
on the choice of orientations of individual faces and hinges. This 
dependence on the orientation should not be present in the final model. 
Furthermore, the weight function should be the same for all faces $F_k$ and hinges $H_m$. To go any further we need to constraint the tensors 
$M_{abc}(F_k)$ and $\Delta^{b_1\cdots b_n}(H_m)$.
That can be done with the help of some algebraic data that we will now introduce.

The first algebraic structure we need is a product on $V$ defined by
\begin{equation}
\phi_a\cdot \phi_b={M_{ab}}^c\phi_c.
\end{equation}
where we are using a tensorial notation with the usual convention
of sum over repeated indices. 
We will use the symbol $A$ to refer to the vector space $V$ to
emphasize that we are now working with an algebra.

In this paper, we will choose $A$ to be the group algebra of $\mathbb{Z}_2$.
The group elements are written as $(-1)^a, a=0,1$ and a basis for $A$
is $\{\phi_0,\phi_1\}$. The product is defined by
\begin{equation}
\phi_a\cdot \phi_b=\phi_{a+b}. 
\label{eq:product}
\end{equation}
Whenever a sum $a+b$ of indices appear, as in (\ref{eq:product}), 
it will always denote sum module 2. This product can also be given in terms 
of the tensor ${M_{ab}}^c$ as
\begin{equation}
{M_{ab}}^c=\delta(a+b,c). 
\label{eq:def-mod-01}
\end{equation}

It is also convenient to define the
dual vector space $A^*$, the dual base $\{\psi^i\}$
and the usual pairing
\begin{equation}
\langle \psi^a,\phi_b \rangle=\delta_b^a.
\end{equation}
We then define the trace $T\in A^*$ as
\begin{equation}
T={M_{ab}}^b\psi^a.
\label{eq:trace}
\end{equation}

Given a face $F_k$, as in figure \ref{fig3}, we
define the associated weight $M_{a_1 a_2 a_3}$ as
\begin{equation}
M_{a_1 a_2 a_3}(z)=\langle T,\phi_{a_1}\cdot \phi_{a_2} \cdot \phi_{a_3}\cdot z\rangle ,
\label{eq:weight-face}
\end{equation} 
where $z=\alpha^0\phi_0+\alpha^1\phi_1$ is a generic element of $A$.
Since the underlining group is abelian, the weight $M_{a_1 a_2 a_3}(z)$ is 
automatically cyclic and gauge invariant. Furthermore, $M_{a_1 a_2 a_3}(z)$ does not
depend on the orientation of $F_k$.

The choice of $z$ in (\ref{eq:weight-face})
will determine the model we are describing. For example, consider the curve 
$(\alpha^0,\alpha^1)=(\frac{1}{2} e^{\beta}, \frac{1}{2} e^{-\beta})$ parametrized by $\beta$.
The corresponding weight
\begin{equation}
M_{a_1 a_2 a_3}(\frac{1}{2} e^{\beta}, \frac{1}{2} e^{-\beta})=e^{\beta (-1)^{a_1 + a_2 + a_3}},
\label{eq:ising-gauge-action}
\end{equation}
corresponds to the spin-gauge model. We can also choose the curve
$(\alpha^0,\alpha^1)=(\frac{1}{2}, \frac{1}{2} e^{-2\beta})$ and that will give
\begin{equation}
M_{a_1 a_2 a_3}(\frac{1}{2}, \frac{1}{2} e^{-2\beta})=e^{-\beta [1-(-1)^{a_1+a_2+a_3}]},
\label{eq:wilson-action}
\end{equation}
describing yet another model.

The second algebraic information is a co-product 
$\Delta:A\rightarrow A\otimes A$ defined by
\begin{equation}
\Delta(\phi_c)={\Delta_c}^{ab}~\phi_a\otimes\phi_b.
\end{equation}
The tensor ${\Delta_c}^{ab}$ defines also a product on the dual space $A^*$.
Using the dual basis $\{\psi^a\}$ we define
\begin{equation}
\psi^a\cdot \psi^b={\Delta_c}^{ab}\psi^c.
\end{equation}
In analogy with (\ref{eq:trace}) and (\ref{eq:weight-face}) we define 
the co-trace $T^*\in A$ as
\begin{equation}
T^*={\Delta_b}^{ba}\phi_a.
\end{equation}
and the tensor $\Delta^{b_1\cdots b_n}$ as
\begin{equation}
\Delta^{b_1\cdots b_n}=
\langle \psi^{a_1}\cdot \psi^{a_2} \cdots \psi^{a_n},T^*\rangle .
\label{eq:weight-hinge}
\end{equation}
It turns out that the co-product that we will need is very simple. We will set
\begin{equation}
{\Delta_c}^{ab}=\delta^a_c\delta^b_c.
\label{eq:def-model-02}
\end{equation}
Therefore 
\begin{equation}
\Delta^{b_1\cdots b_n}=\left\lbrace 
\begin{tabular}{l}
1 \mbox{ if all indices are equal}\\
0 \mbox{ otherwise}
\end{tabular}\right. . 
\end{equation}
Note that the orientations of hinges do not affect the corresponding
statistical weight.

The tensors ${M_{ab}}^c$ and ${\Delta_a}^{bc}$ given in (\ref{eq:def-mod-01}) and 
(\ref{eq:def-model-02}) together with the weights $M_{abc}$ and $\Delta^{a_1\cdots a_n}$ 
defined by (\ref{eq:weight-face}) 
and (\ref{eq:weight-hinge}) completely specify our model. 
It is a simple exercise to show that the partition function (\ref{eq:big-model}) 
reduces to
\begin{equation}
Z(\alpha^0,\alpha^1)= \sum_{\{\sigma_l\}} 
                      \prod_{f\in T(\mathcal{M})} M_{a_1 a_2 a_3}(f),
\label{eq:spin-gauge}
\end{equation}
where the sum $\{\sigma_l\}$ is over the configurations on the links $l\in T(\mathcal{M})$ and
$M_{abc}(f)$ is the weight of a configuration $(a_1, a_2, a_3)$ on the face $f\in T(\mathcal{M})$.
Notice that 
\begin{equation}
Z(\lambda \alpha^0,\lambda\alpha^1)=\lambda^{N_F}Z(\alpha^0,\alpha^1).
\label{eq:Z-scale}
\end{equation}

As $M_{a_1 a_2 a_3}$
depends on $z=\alpha^0\phi_0+\alpha^1\phi_1\in A$, the model 
depends on two parameters $(\alpha^0,\alpha^1)$. We will denote the
parameter space by $\Gamma$ .
The parameters $(\alpha^0,\alpha^1)$ are related to the parameters $(m_0,m_1)$ of (\ref{eq:theModel})
as
\begin{eqnarray}
m_0 &=&\alpha^0 +\alpha^1 \nonumber \\
m_1 &= & \alpha^0 -\alpha^1.
\end{eqnarray}

Let us recall that
the algebra defined in (\ref{eq:product}) is a group algebra and as such it is
also a Hopf algebra with co-product coming from ${\Delta_a}^{bc}$. The maps antipode 
$S:A\rightarrow A$, unity $e:\mathbb{C}\rightarrow A$ and
co-unity $\epsilon:A\rightarrow \mathbb{C}$ can be described in terms of tensors as 
$S(\phi_a)=S_a^b\phi_b$, $e(1)=e^a\phi_a$  and $\epsilon(\phi_a)=\epsilon_a$.
For the case of $\mathbb{Z}_2$ we have
\begin{equation}
S_a^b=\delta_a^b,~~ e_a=\delta_a^0 ~\mbox{ and }~ \epsilon_a=1.
\end{equation}
Notice that these tensors are essentially trivial and will not 
show up explicitly in the calculations. For a non abelian case,
for example, $S_a^b$ is related to the orientation but that will play
no role in the $\mathbb{Z}_2$ case.

\section{Quasi-Topological Limits} \label{sec:limits}

\begin{figure}[h!]
	\begin{center}
		\includegraphics[scale=.8]{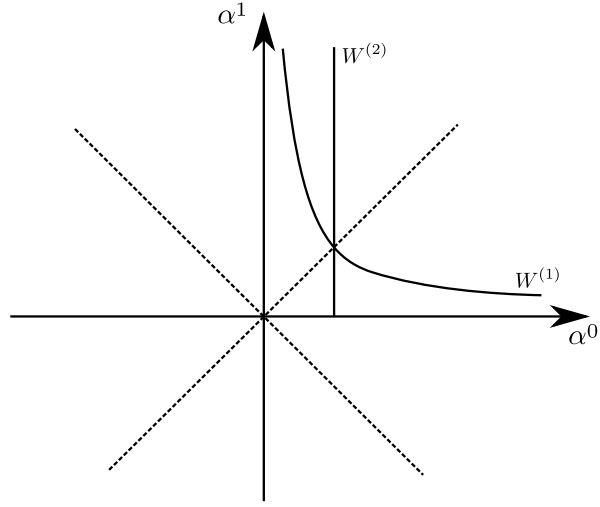}
		\caption{The set $\Gamma_0$ consists of the two axes plus the dashed diagonal lines. The models $W^{(1)}$ and $W^{(2)}$ are curves on $\Gamma$.}
		\label{fig4}
	\end{center}
\end{figure}

We would like to explore the model (\ref{eq:spin-gauge}) %{\tiny eq:spin-gauge}
and look for points of the parameter space $(\alpha^0,\alpha^1)$ where the
model has a topological or quasi-topological behaviour. The simplest case is the
one considered in \cite{kuperberg,cfs}. It corresponds the point 
$(\alpha^0,\alpha^1)=(1,0)$ in the parameter space $\Gamma$ or, equivalently, to the 
choice in (\ref{eq:weight-face}) of $z$ equals to the identity $\phi_0$ of the algebra.
For this particular point of the parameter space we can 
bring in the Hopf structure of $A$, follow
\cite{kuperberg,cfs} and conclude that 
\begin{equation}
Z\left(T(\mathcal{M}),\alpha^0=1,\alpha^1=0\right)= 2^{N_F + N_L - N_T} Z^{top} \left( \mathcal{M}\right),
\label{eq:Z(1,0)}
\end{equation}
where $N_T$, $N_F$ and $N_L$ are the number of tetrahedra, faces and links of $T(\mathcal{M})$.
However,  $(\alpha^0,\alpha^1)=(1,0)$ is not the only quasi-topological point.
In this section we will show that there is a sub-set $\Gamma_0\subset \Gamma$ 
with dimension one such that the partition function is quasi-topological.
In figure \ref{fig4} we have the parameter space $\Gamma$ where the set $\Gamma_0$ 
consists of four straight lines: the diagonals and the axis. We also have included the models with weights
$W^{(1)}$ and $W^{(2)}$ defined in (\ref{eq:spin}) and (\ref{eq:wilson}).
They are curves parametrized by a single parameter $\beta $.

Let us consider $(\alpha^0,\alpha^1)=(\lambda,0)$. 
It corresponds to $z=\lambda \phi_0\in A$. The new weight is simple
$M_{abc}(z)=\lambda M_{abc}(\phi_0)$. Therefore the partition function is
the same as for $(\alpha^0,\alpha^1)=(1,0)$ multiplied by a factor. In other
words
\begin{equation}
Z(T(\mathcal{M}),\lambda,0)=\lambda^{N_F} 2^{N_F + N_L - N_T} Z^{top} \left( \mathcal{M}\right).
\label{eq:Z(lambda,0)}
\end{equation}

The tensor components $M_{abc}(z)$ and $\Delta^{b_1\cdots b_n}$ depend on the choice of  a basis of the algebra 
$A$. Therefore, different choices of basis will lead to different weights and
therefore different models.
However, the partition function  
$Z(T(\mathcal{M}),\alpha^0,\alpha^1)$ has been written as a scalar
and therefore is invariant under a change of basis. This large invariance of 
$Z(T(\mathcal{M}),\alpha^0,\alpha^1)$ has been interpreted as dualities between different models. This fact has been explored by us in 
\cite{n2} to show that the classical Kramers and Wannier duality relations
are special cases of these more general dualities.
In what follows we will show that there is a duality relation between the model
at $(\alpha^0,\alpha^1)=(\lambda,0)$ and $(\alpha^0,\alpha^1)=(0,\lambda)$.
We will show that 
\begin{equation}
Z\left(T(\mathcal{M}),0,\lambda\right)=Z(T(\mathcal{M}),\lambda,0)~,
\label{eq:duality}
\end{equation}
which allow us to compute $Z(T(\mathcal{M}),0,\lambda)$.  

Let us start by recalling that at the topological point 
$(\alpha^0,\alpha^1)=(1,0)$ the weights read
\begin{eqnarray}
M_{abc}(\phi_0)&=&2\delta(a+b+c,0)~,\\
\Delta^{b_1\cdots b_n}&=& \delta(b_1,b_n)\delta(b_2,b_n)\cdots \delta(b_{n-1},b_n).
\end{eqnarray}
Consider the matrix
\begin{equation}
E=\left(
\begin{tabular}{cc}
0 & 1\\
1 & 0
\end{tabular}
\right)~,
\end{equation}
and a new basis $\{\phi^\prime_0,\phi^\prime_1\}$ defined as 
$\phi^\prime_a=E_a^b\phi_b$. This transformation simply changes
an index $a$ by $\bar a$, where $\bar 0=1$ and $\bar 1=0$. In the new basis
we have
\begin{eqnarray}
M^\prime_{abc}(1,0)&=&M_{\bar{a}\bar{b}\bar{c}}(1,0)= 
2\delta(\bar{a}+\bar{b}+\bar{c},0)=2\delta(a+b+c,1)~,\\
\Delta^{\prime b_1\cdots b_n}&=& 
\delta(\bar b_1,\bar b_n)\delta(\bar b_2,\bar b_n)\cdots \delta(\bar b_{n-1},\bar b_n)~,
\nonumber \\
&=&\delta(b_1,b_n)\delta(b_2,b_n)\cdots \delta(b_{n-1},b_n)~.
\end{eqnarray}
Note that  
\begin{equation}
\Delta^{\prime b_1\cdots b_n}=\Delta^{b_1\cdots b_n}.
\label{eq:iqualdade-D}
\end{equation}
On the other hand, the tensor components $M_{abc}(0,1)$ on the original basis 
is $M_{abc}(0,1)=\delta(a+b+c,1)$. Therefore 
\begin{equation}
M^\prime_{abc}(1,0)=M_{abc}(0,1).
\label{eq:iqualdade-M}
\end{equation}
Equations  (\ref{eq:iqualdade-D}) and (\ref{eq:iqualdade-M}) imply
that $Z(T(\mathcal{M}),0,\lambda)=Z(T(\mathcal{M}),\lambda,0)$ as announced.

The partition function corresponding to the diagonal line $\alpha^0=\alpha^1$ 
in figure \ref{fig4} can be obtained by choosing $z_{int}=\phi_0+\phi_1$ in (\ref{eq:weight-face}). 
One can see that $z_{int}$ is such that $\phi_g \cdot z_{int}=z_{int}$. 
In a Hopf algebra, such element is called a co-integral \cite{hopf}. 
Therefore
\begin{equation}
M_{abc}(1,1)=\langle T,z_{int}\rangle=2.
\end{equation} 
The fact that the weight $M_{abc}(1,1)$ is independent of the configurations turns the
computation of $Z(T(\mathcal{M}),\lambda,\lambda)$ 
completely trivial. One can immediately see that
\begin{equation}
Z(T(\mathcal{M}),\lambda,\lambda)=2^{N_F+N_L}\lambda^{N_F}.
\end{equation}
Note that $Z(T(\mathcal{M}),\lambda,\lambda)$ 
depends only on the size of the lattice $T(\mathcal{M})$. 
In contrast with (\ref{eq:Z(lambda,0)}), the topological invariant is trivial since
the partition function does not to depend on the topology of the underlining manifold 
$\mathcal{M}$.

To compute the partition function for  $\alpha^1=-\alpha^0$ it is enough to consider
the point $(\alpha^0,\alpha^1)=(1,-1)$. That is the same as setting $z=\phi_0-\phi_1$
on (\ref{eq:weight-face}). The weight $M_{abc}(1,-1)$ reads
\begin{equation}
M_{abc}(1,-1)=2(-1)^{a+b+c}.
\end{equation}
Instead of computing $Z(T(\mathcal{M}),1,-1)$ directly, we will make use of the duality
related to change of basis. Let us choose another basis by applying the transformation
matrix $E_a^b=(-1)^a\delta_a^b$. In the new basis the weights are
\begin{eqnarray}
M^\prime_{abc}&=&2~, \nonumber \\
(\Delta ^{\prime})^{a_1a_2\cdots a_n}&=&(-1)^{a_1+a_2+\cdots +a_n} 
           \delta_{a_n}^{a_1}\delta_{a_n}^{a_2}\cdots\delta_{a_n}^{a_{n-1}}~.
\label{eq:dual-model}
\end{eqnarray}
The resulting model has weights associated to the links only and the partition function
can be easily computed. After plugging (\ref{eq:dual-model}) in (\ref{eq:big-model}) and 
taking into account the scaling factor we have
\begin{equation}
Z(T(\mathcal{M}),\lambda, -\lambda)=\lambda^{N_F}\prod_{l}(1+(-1)^{I(l)})~,
\label{eq:(1,-1)}
\end{equation}
where the product runs over the links of the lattice and $I(l)$ denotes 
the number of faces that share the link $l$. Note that $I(l)$ is not
of topological nature. Furthermore, the partition function vanishes 
whenever there is a link that is shared by a odd number of faces.
This is an indication that $(\alpha^0, \alpha^1)=(\lambda, -\lambda)$ 
is a very peculiar model.

%There is always the possibility of taking regular 
%lattices with $I(l)$ even for all links $l\in L$. 
%usual regular cubic lattice.

It is clear from the computation of the partition function 
that the lines that make up $\Gamma_0$ are not all equivalent.
Actually they are all different from each other. It is true that
$Z(T(\mathcal{M}),\lambda,0)$ is the same as $Z(T(\mathcal{M}),0,\lambda)$. 
However, the expectation value of Wilson loops are not the same for these
two models. As we will show in the next section, only 
$(\lambda,0)$ is in fact a topological theory.

Before we conclude this section, we would like to point out the
relation between $\Gamma_0$ and familiar models such as the ones defined by
(\ref{eq:spin}) (spin-gauge model) and (\ref{eq:wilson}) 
(gauge theories). As we have discussed before, the spin-gauge model 
given in (\ref{eq:ising-gauge-action})
corresponds to the curve $(\alpha^0,\alpha^1)=(1/_2 e^\beta,1/_2 e^{-\beta})$. It is clear 
from figure \ref{fig4} that this curve  approaches  $\Gamma_0$ as $\beta$ goes to
$+\infty$ and $-\infty$. Another point of contact with $\Gamma_0$
is when $\beta\rightarrow 0$. As for the model (\ref{eq:wilson}), only the
limits $\beta\rightarrow 0$ and $\beta\rightarrow \pm \infty$ are part of
$\Gamma_0$.
 
\section{Wilson Loops}\label{sec:wilsonLoops}

The two parameter gauge model (\ref{eq:big-model-simple}) of last section  have 
numerical quantities $\langle W_R(\gamma )\rangle$ 
that 
are the natural generalization of the expectation value of Wilson loops
for a closed curve $\gamma $ and irreducible
representation $R$. The definition of $\langle W_R(\gamma )\rangle$ reduces to the 
familiar expression when restricted to the usual gauge models. 
In this section we will define  and compute these observables for $\Gamma_0$. 

For simplicity, we will start by considering the loop $\gamma$ to be unknotted. Knotted loops will be considered in the last part of this section.

Let $\gamma$ be a loop made of a set of links  
$\omega(\gamma)=\{\omega_1,...,\omega_p\}$. For such a loop
we introduce the tensor $W_{a_1 a_2 \cdots a_p }$ with $p$  indices given by
\begin{equation}
W^R_{a_1\cdots a_p}=\langle T,\phi_{a_1}\cdots \phi_{a_p}\cdot z_R\rangle,
\label{eq:WL-weight-1} 
\end{equation}
where $a_k$ is the configuration at link $l_k$ and $z_R$ is the unique element in the center of $A$ such that
\begin{equation}
\langle T,\phi_g\cdot z_R\rangle= \mbox{Tr}_R(g),
\end{equation} 
where $R$ denotes an irreducible representation of the group.
 
The group $\mathbb{Z}_2$ has only two irreducible representations labelled $R=0$ and $R=1$, such that

\begin{eqnarray}
\mbox{Tr}_0(\phi_a)&=&1~, \nonumber \\ 
\mbox{Tr}_1(\phi_a)&=&(-1)^a~. \nonumber
\end{eqnarray}
We only need to consider the non-trivial representation.
In other words, we will set $z_R=\frac{1}{2}\left(\phi_0-\phi_1\right)$ since
that will give us $\langle T,\phi_a\cdot z_R\rangle=(-1)^a$. 
For now on we will omit the index $R$ indicating the 
representation simply write (\ref{eq:WL-weight-1}) as 
\begin{equation}
W_{a_1\cdots a_p}=(-1)^{a_1+\cdots +a_p}.
\label{eq:WL-weight-2} 
\end{equation}

\begin{figure}[h!]
	\begin{center}
		\includegraphics[scale=1]{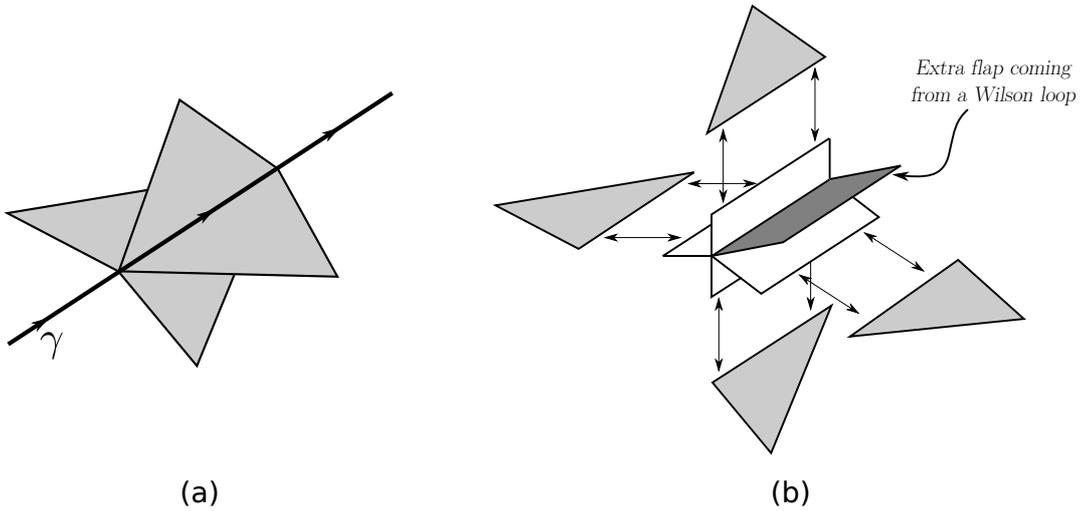}
		\caption{A small region of the lattice with a Wilson loop $\gamma$ is illustrated in {\bf (a)}. Figure {\bf (b)} shows the corresponding decomposition. For a link of $\gamma$, the corresponding hinge has an extra flap as indicated.}
		\label{fig5}
	\end{center}
\end{figure}

\begin{figure}[h!]
	\begin{center}
		\includegraphics[scale=1]{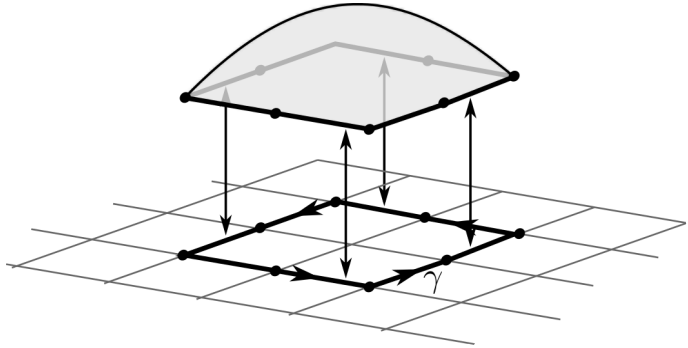}
		\caption{Wilson loops can be thought as an extra face which is not part of the lattice.}
		\label{fig6}
	\end{center}
\end{figure}

We would like to construct $\langle W(\gamma)\rangle$ as a scalar
in the same way it has been done for the partition function 
(\ref{eq:big-model-simple}). As before, we make use of the contra-variant
tensors $\Delta^{b_1\cdots b_n}$ associated to the links. Consider
a link $l$ shared by $I(l)$ faces.
If $l$ does not belong to the loop $\gamma$, the corresponding tensor is
the same as for the partition function and will be written as 
$\Delta^{b_1\cdots b_{I(l)}}$. If, however, the link in question is 
 one of the links of $\gamma$ ($l=\omega_j\in \omega(\gamma))$, the corresponding
tensor will be $\Delta^{b_1\cdots b_{I(l)}a_j}$. Note that the new tensor has
an extra contra-variant index $a_j$.
The expectation value of the Wilson loop is defined to be
\begin{equation}
\langle W(\gamma)\rangle=\frac{1}{Z}
\prod_{f\in \{F\}}  M_{abc}(f) \prod_{l\notin \omega(\gamma) } \Delta^{b_1\cdots b_{I(l)}}(l)
\prod_{\omega_j\in \omega(\gamma) } \Delta^{b_1\cdots b_{I(l)}a_j}
W_{a_1\cdots a_p},
\label{eq:wilson-loop}
\end{equation}
where we are using the simplified notation of (\ref{eq:big-model-simple}). The contraction
of indices are as follows:
each covariant index of $W_{a_1\cdots a_p}$ is to be contracted with the
extra index $a_j$ in $\Delta^{b_1\cdots b_{I(l)}a_j}$ as 
explicitly indicated. The indexes from the product 
$\prod _{f\in \{F\}} M_{abc}$ follows the same rule as for the partition function
when contracting with the $b_k$ indexes in 
$\Delta^{b_1\cdots b_{I(l)}}$ and $\Delta^{b_1\cdots b_{I(l)}a_j}$.
An interpretation of (\ref{eq:wilson-loop}) in terms of 
gluing of hinges $(H_m)$ and faces $(F_k)$ that is analogue to
the partition function can also be given. To each link $l$ we associate 
a hinge $H_l$. If the link is not part of the loop $\gamma$, the 
corresponding hinge has exactly $I(l)$ flaps that will connect to
$I(l)$ faces in the usual way (see figure \ref{fig3}). For links 
$\omega_j\in \omega(\gamma)$ the corresponding hinge $H_{\omega_j}$ has 
$\left( I(\omega_j)+1 \right)$ flaps. After gluing the $I(\omega_j)$ faces to
$H_{\omega_j}$, we are left with extra flaps, one for each link
of $\omega (\gamma )$ as illustrated in figure \ref{fig5}. The contraction of
indices $a_j$ in (\ref{eq:wilson-loop}) can be seen as the attachment
of a polyhedral face $W_\gamma$ with $p$ edges. 
An example of such attachment is shown in figure \ref{fig6}.
This special face
$W_\gamma$ has weight $W_{a_1\cdots a_p}$ and is not to be thought
as part of the lattice. In general, it will not be possible
to embed $W_\gamma$ in $3D$ space.

The expectation value of the Wilson loop will be a function  
$\langle W(\gamma)\rangle(\alpha^0,\alpha^1)$. As a consequence of
(\ref{eq:Z-scale}) one can see that
\begin{equation}
\langle W(\gamma)\rangle(\lambda\alpha^0,\lambda\alpha^1)=
\langle W(\gamma)\rangle(\alpha^0,\alpha^1).
\label{eq:W-scale}
\end{equation}
Therefore we only need to compute $\langle W(\gamma)\rangle(\alpha^0,\alpha^1)$
for one point on each straight line of $\Gamma_0$ .

We now proceed with the computation of $\langle W(\gamma)\rangle$
when the parameters $(\alpha^0,\alpha^1)$ belongs to $\Gamma_0$. That
can be divided in three cases as follows.
 
\mbox{ }

\noindent
{\it \bf Case 1} ($z=\phi_0+\phi_1$):\\
The first case correspond to $(\alpha^0,\alpha^1)=(1,1)$. As we have
seen, $M_{abc}=2$. All the sums on $b$ indices in (\ref{eq:wilson-loop})
are straightforward due to the Kronecker deltas in $\Delta^{b_1\cdots b_{I(l)}}$
and $\Delta^{b_1\cdots b_{I(l)}a_j}$. We are then left with
\begin{equation}
\langle W(\gamma)\rangle \propto \prod_{i=1}^p\sum_{a_i}(-1)^{a_i}=0~.
\end{equation}
This result does not depend on the loop $\gamma$.

\mbox{ }

\noindent
{\it \bf Case 2} ($z=\phi_0-\phi_1$):\\

For this particular case we have $W_{a_1\cdots a_p}=(-1)^{a_1+\cdots +a_p}$.
Notice that this is the same function as
the weights $M_{abc}=2(-1)^{a+b+c}$ for the faces. Therefore the 
numerator in (\ref{eq:wilson-loop}) is the same thing as a
partition function with an extra face determined by the loop $\gamma$
(see figure \ref{fig6}). Using (\ref{eq:(1,-1)}) we can write
\begin{equation}
\langle W(\gamma)\rangle=\frac{1}
{\prod_{l}[1+(-1)^{I(l)}]}\prod_{l\notin \omega(\gamma)}[1+(-1)^{I(l)}]
\prod_{l\in \omega(\gamma)}[1+(-1)^{I(l)+1}]~.
\end{equation}

This region of the parameter space is quite peculiar. 
Notice that the denominator of $\langle W(\gamma)\rangle$ 
vanish if $I(l)$ is odd for some $l$ and $\langle W(\gamma)\rangle$
is not well defined. When $I(l)$ is even for all links, $\langle W(\gamma)\rangle$
is well defined but it is equal to zero due to the factors $[1+(-1)^{I(l)+1}]$.
 
\mbox{ }

\noindent
{\it \bf Case 3} ($z=\phi_g, g=0,1$):\\

The points with coordinates $(\alpha^0,\alpha^1)$ given by $(1,0)$ and $(0,1)$ in $\Gamma_0$ 
correspond to $z=\phi_g$ with $g=0$  and $g=1$ respectively.
We know from previews sections that the partition function is the same for these two cases.
However, this is not true for $\langle W(\gamma)\rangle$.

\begin{figure}[h!]
	\begin{center}
		\includegraphics[scale=1]{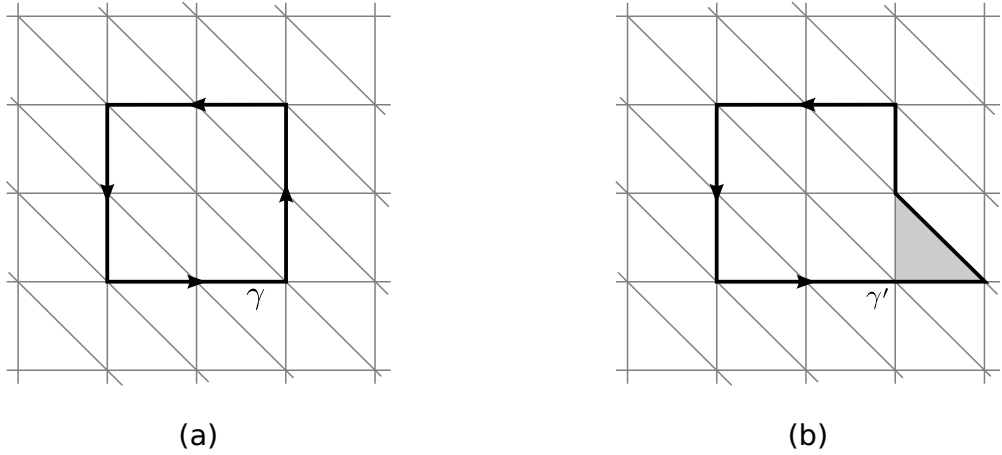}
		\caption{The loop $\gamma^\prime$ in {\bf (b)} is a small deformation of the loop $\gamma$ in {\bf (a)}.}
		\label{fig7}
	\end{center}
\end{figure}

We will show that $\langle W(\gamma)\rangle$ is quasi-topological in the sense
that it does not depend strongly on the geometry of $\gamma$. The idea is
to investigate the behaviour of  $\langle W(\gamma)\rangle$ under small deformations.
In a triangulation it is natural to define a small deformation of a loop $\gamma$
as follows. Let $\omega(\gamma)$ be  the set of links of 
$\gamma$. A small deformation of $\gamma$ is a local move that replaces one
link $\omega_1\in \omega(\gamma$) by a pair of links 
$\omega_2,\omega_3$ such that 
$\omega_1,\omega_2$ and $\omega_3$ belong to the same face. 
There is also the reverse move where a pair of links $\omega_1,\omega_2$ is
replaced by $\omega_3$ provided they belong to the same face.
A small deformation is shown in figure \ref{fig7}. 
With this definition of a small deformation we now establish how
$\langle W(\gamma)\rangle$ changes under small deformations for $z=\phi_g, g=0,1$. 

\begin{figure}[h!]
	\begin{center}
		\includegraphics[scale=1]{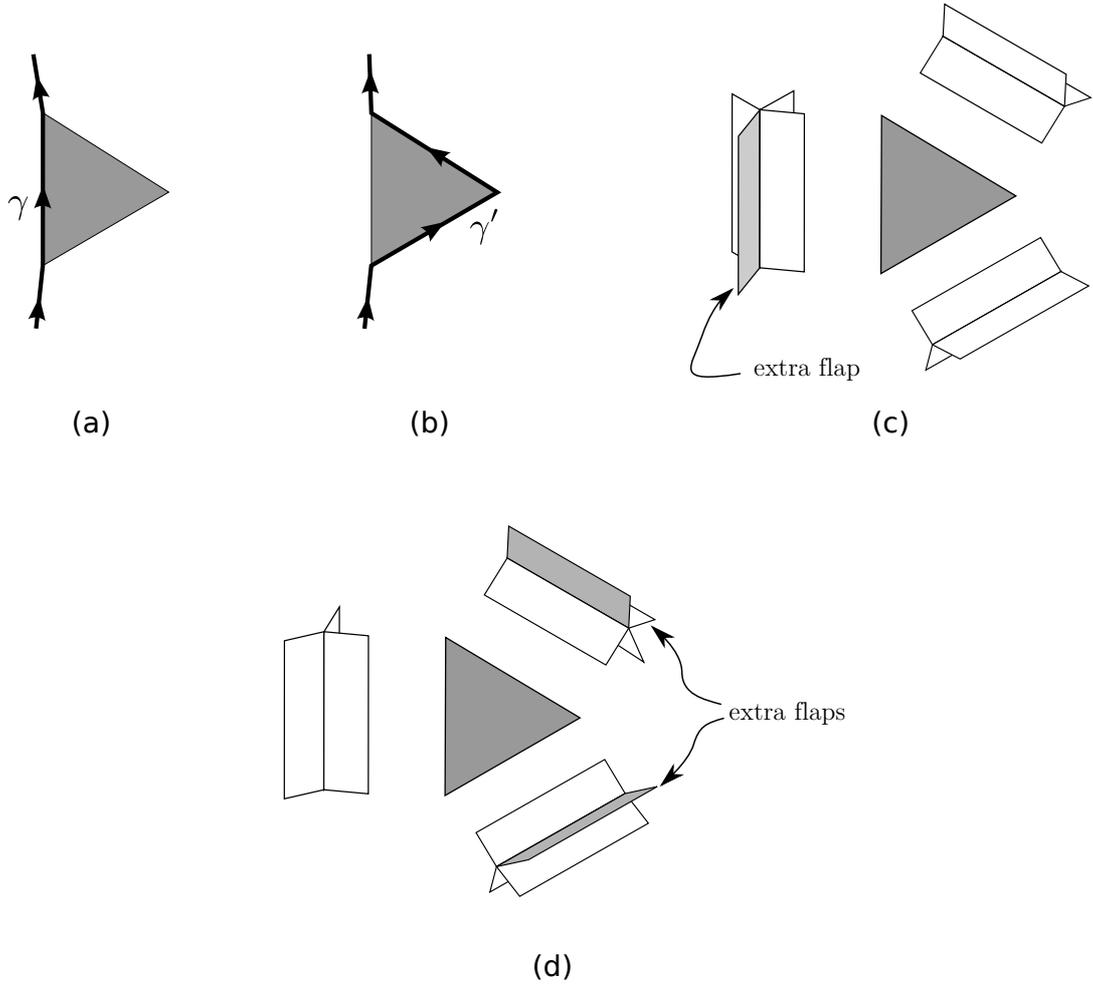}
		\caption{In {\bf (a)}-{\bf (b)} we represent a small deformation of a loop $\gamma$ in a lattice and in {\bf (c)}-{\bf (d)} we represent the same deformation in a lattice decomposition.}
		\label{fig8}
	\end{center}
\end{figure}

Consider figure \ref{fig8} {\bf (a)} where we have single out a small part of the partition 
$(F_k,H_j)$. Only the elements connected to the link $\omega_1$ are relevant.
In figure \ref{fig8} {\bf (b)} we have performed a small deformation of $\gamma$ by replacing
$\omega_2,\omega_3$ by $\omega_1$. The weights associated to figure \ref{fig8} {\bf (c)} 
and figure \ref{fig8} {\bf (d)} are the tensors $A$ and $B$ given by
\begin{eqnarray}
A^{a_2\cdots a_pb_2\cdots b_qc_2\cdots c_r}_{\alpha_1\cdots\alpha_s}
=M_{a_1b_1c_1}\Delta^{a_1\cdots a_p\omega_1} 
   \Delta^{b_1\cdots b_q} \Delta^{c_1\cdots c_r}
   W_{\omega_1\alpha_1\cdots \alpha_s}\nonumber \\
=\sum _{a_1,b_1,c_1}M_{a_1b_1c_1}\Delta^{a_1\cdots a_p} 
   \Delta^{b_1\cdots b_q} \Delta^{c_1\cdots c_r}W_{a_1\alpha_1\cdots \alpha_s}
\label{eq:A}
\end{eqnarray}   
and
\begin{eqnarray}
B^{a_2\cdots a_pb_2\cdots b_qc_2\cdots c_r}_{\alpha_1\cdots\alpha_s}
=M_{a_1b_1c_1}\Delta^{a_1\cdots a_p} 
   \Delta^{b_1\cdots b_q\omega_2} \Delta^{c_1\cdots c_r\omega_3}
   W_{\omega_2 \omega_3\alpha_1\cdots \alpha_s}\nonumber \\
=\sum _{a_1,b_1,c_1}M_{a_1b_1c_1}\Delta^{a_1\cdots a_p} 
   \Delta^{b_1\cdots b_q} \Delta^{c_1\cdots c_r}
   W_{b_1 c_1\alpha_1\cdots \alpha_s},
\label{eq:B}
\end{eqnarray} 

\noindent
where we have performed the sums on $\omega_1,\omega_2$ and $\omega_3$. We also have
explicitly written the sums on $a_1,b_1$ and $c_1$. Note that 
$W_{a_1\alpha_1\cdots \alpha_s}=(-1)^{a_1}W_{\alpha_1\cdots \alpha_s}$,
$W_{b_1c_1\alpha_1\cdots \alpha_s}=(-1)^{b_1+c_1}W_{\alpha_1\cdots \alpha_s}$
and $M_{a_1b_1c_1}=2\delta(a_1+b_1+c_1,g)$. That is enough to show that
\begin{equation}
M_{a_1b_1c_1}W_{b_1 c_1\alpha_1\cdots \alpha_s}=
(-1)^gM_{a_1b_1c_1}W_{a_1\alpha_1\cdots \alpha_s},
\label{eq:MW=MW}
\end{equation}
therefore
\begin{equation}
A^{a_2\cdots a_pb_2\cdots b_qc_2\cdots c_r}_{\alpha_1\cdots\alpha_s}=
(-1)^gB^{a_2\cdots a_pb_2\cdots b_qc_2\cdots c_r}_{\alpha_1\cdots\alpha_s}.
\label{eq:A=B}
\end{equation}

We see that $\langle W(\gamma)\rangle$ is invariant under small deformations 
for $g=0$ that correspond to $(\alpha^0,\alpha^1)=(1,0)$. As for $g=1$ or 
$(\alpha^0,\alpha^1)=(0,1)$, 
the number $\langle W(\gamma)\rangle$ flips
sign each time we perform a small deformation on $\gamma$.

The simplest loop $\gamma_0$ is made of a single triangular face with links
$\omega(\gamma)=\{\omega_1,\omega_2,\omega_3\}$.
If we recall that $M_{abc}=
2 \delta(a+b+c,g)$ and 
$W_{\omega_1\omega_2\omega_3}=(-1)^{\omega_1+\omega_2+\omega_3}$ ,
it becomes a straightforward computation to show that
$
\langle W(\gamma_0)\rangle=(-1)^g.
$
We can now deform the smallest loop $\gamma_0$ by adding $(N-1)$ triangles and
arriving at a planar loop $\gamma_N$. For such a loop we get
\begin{equation}
\langle W(\gamma_N)\rangle=(-1)^{gN}; ~~g=0,1.
\label{eq:area-law}
\end{equation} 

Equation (\ref{eq:area-law}) for $g=1$ shows 
$\langle W(\gamma_N)\rangle$ as a function of the number of triangles swept
in the process of stretching $\gamma_0$ into $\gamma_N$. This function
is a very simple "area law". It depends only on the parity of $N$. Notice that each 
time we add a triangle to $\gamma_0$, the
number of links of the loop also changes by one unity. Therefore we could
have written 
\begin{equation}
\langle W(\gamma_N)\rangle=(-1)^{gN_L}~,
\label{eq:perimeter}
\end{equation}
where $N_L$ is the number of links of the loop $\gamma$. We could interpret this 
formula as a "perimeter law". The fact that there is no distinction
between area or perimeter law is peculiar to the gauge group $\mathbb{Z}_2$
and the fact that we are using triangular lattices. For square lattices
the variation in the number of links is even and (\ref{eq:perimeter}) does
not hold, but (\ref{eq:area-law}) is still true.

So far we have considered only planar loops. For knotted loops, the expectation value $\langle W(\gamma_N)\rangle$ for $g=0$ may
depends on the class of isotopy of the  loop $\gamma$. Equation
(\ref{eq:area-law}) is valid for loops $\gamma$ that can be deformed to
the trivial knot. If $\gamma$ is a non trivial knot, equation (\ref{eq:area-law})
may not hold. 
We will show that $\langle W(\gamma_N)\rangle$ actually 
does not depend on the class of isotopy of $\gamma$ and therefore 
(\ref{eq:area-law}) is still correct.

\begin{figure}[h!]
	\begin{center}
		\includegraphics[scale=1]{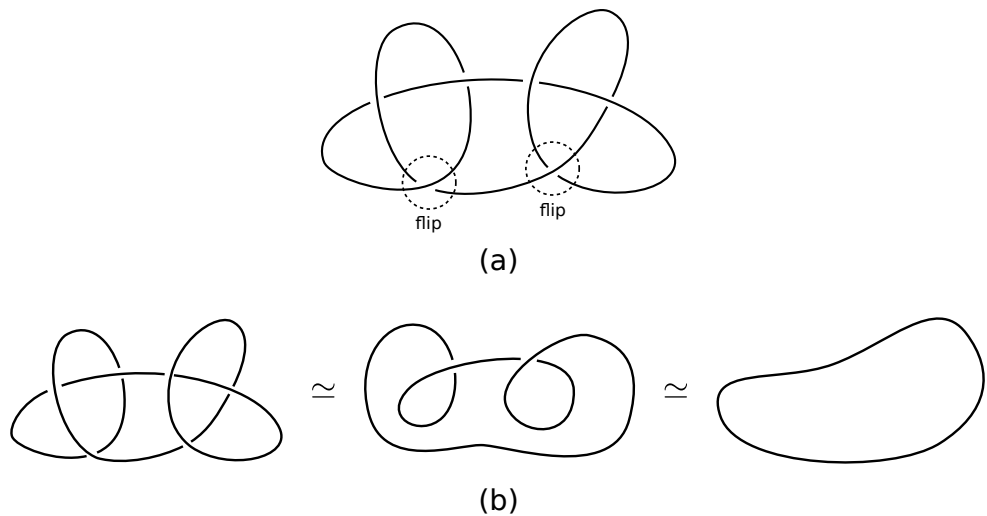}
		\caption{In {\bf (a)} we have a knotted loop that can be made trivial by two flip moves, as illustrated in {\bf (b)}.}
		\label{fig9}
	\end{center}
\end{figure}

Let $\gamma$ be a knotted loop and
its knot diagram as illustrated in figure \ref{fig9} {\bf (a)}. We can produce a
new knot diagram by flipping under-crossing into over-crossings and vice versa. 
This flips are local moves that only affect the knot in a small region.
It is a well known result from knot theory that any $\gamma$ can be made into
the trivial knot if we perform a number  of flips, as we can see in figure \ref{fig9} {\bf (b)}. We will show that
$\langle W(\gamma_N)\rangle$ is invariant by flips and therefore does 
not depend on the isotopy class of $\gamma$.

\begin{figure}[h!]
	\begin{center}
		\includegraphics[scale=1]{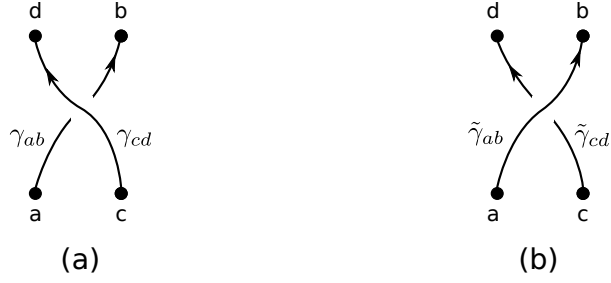}
		\caption{The situation shown in {\bf (a)} represents a flip move of one shown in {\bf (b)} and vice versa. }
		\label{fig10}
	\end{center}
\end{figure}

\begin{figure}[h!]
	\begin{center}
		\includegraphics[scale=1]{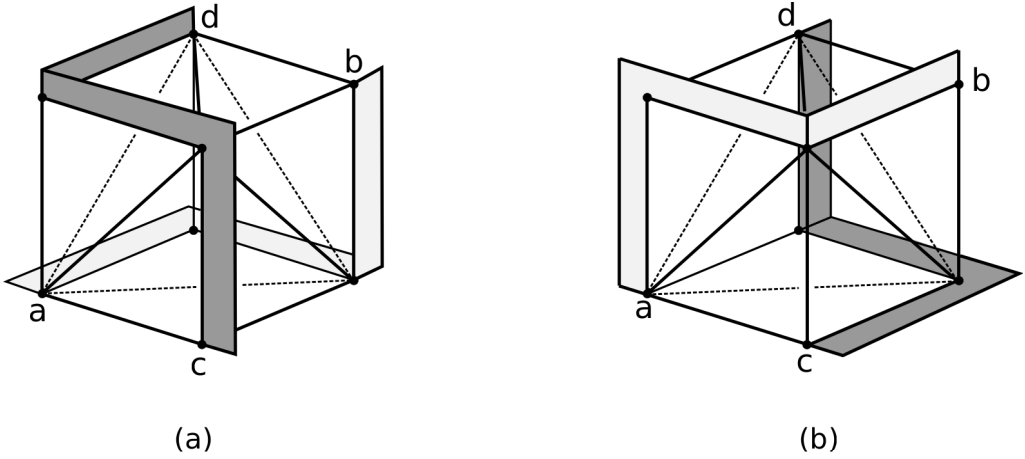}
		\caption{Here we can see how a flip move acts on curves in a surface $\partial B = S^2$ of a ball.}
		\label{fig11}
	\end{center}
\end{figure}

Let us consider a 3-dimensional ball ${B}$ around a crossing in a knot $K$. That
will give us  curves $\gamma_{ab}$ and $\gamma_{cd}$ connecting 
the points $(a,b)$ and $(c,d)$ at the surface $\partial B=S^2$ of $B$ as in figure \ref{fig10} {\bf (a)}.
After a flip move, we have a new knot $\tilde K$ and  new curves are 
$\tilde\gamma_{ab}$ and $\tilde\gamma_{cd}$ as
illustrated in \ref{fig10} {\bf (b)}. Before analysing the general case, we will look at a 
simple example where $B$ is a cube inside the triangulation and the curves, before
and after the flip move, are the ones given in figure \ref{fig11} {\bf (a)} and {\bf (b)}.
In this figures we are using the interpretation of the expectation value of the Wilson loop as an extra flap 
as was explained in section \ref{sec:wilsonLoops}.

\begin{figure}[h!]
	\begin{center}
		\includegraphics[scale=1]{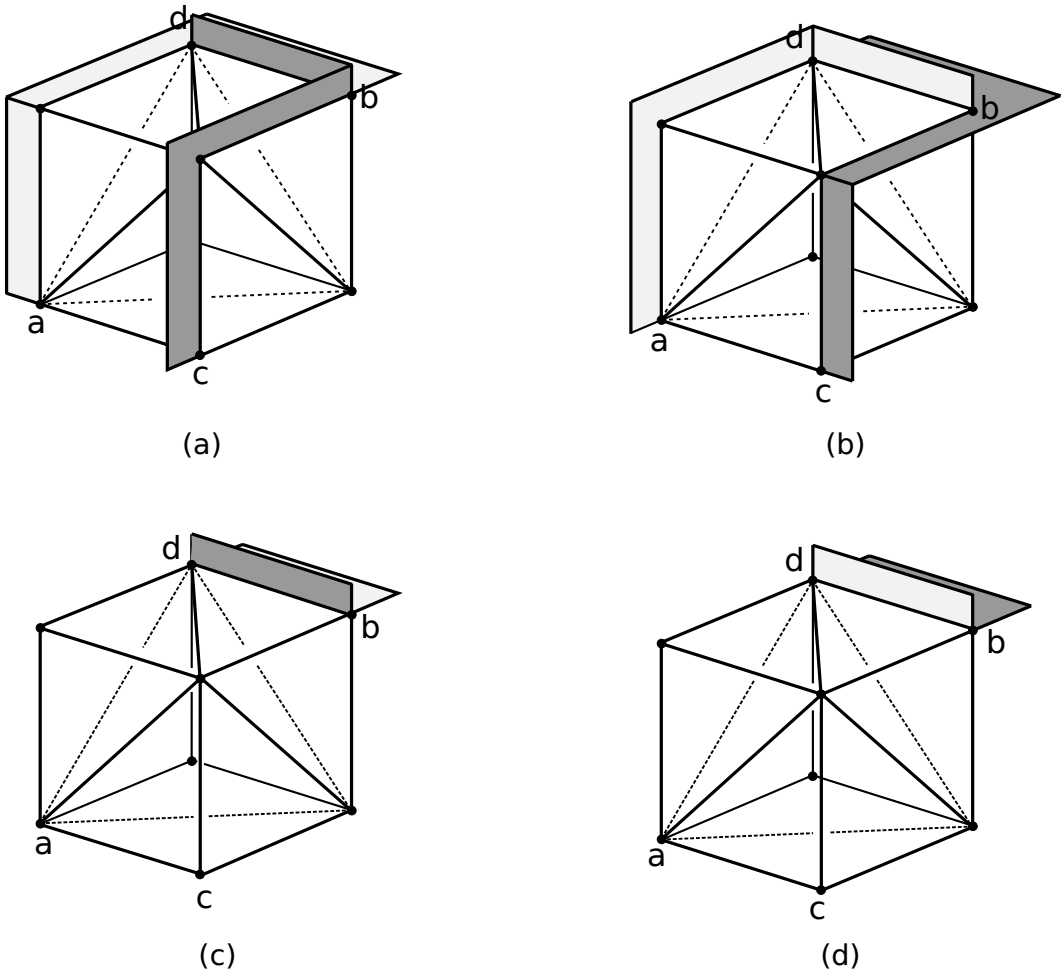}
		\caption{The pictures {\bf (a)} and {\bf (b)} differs from each other only by the link $(d,b)$, as we can see in {\bf (c)} and {\bf (d)}.}
		\label{fig12}
	\end{center}
\end{figure}

One can see that the the sequence of extra flaps can be deformed into the sequences given by figure \ref{fig12} {\bf (a)} and {\bf (b)}. After we perform these deformations, 
the computation of  $\langle W(K)\rangle$ and of $\langle W(\tilde K)\rangle$
differ only at a single link. The difference is that the flaps coming from the two curves
are swaped. In one case, the curves will contribute to (\ref{eq:wilson-loop}) 
with a factor
\begin{equation}
\Delta^{a_1\cdots a_n\omega_1\omega_2}W^{(1)}_{\omega_1\alpha_1\cdots\alpha_s}
   W^{(2)}_{\omega_2\beta_1\cdots\beta_r}~,
\end{equation}
  and in the other case the factor is
\begin{equation}
\Delta^{a_1\cdots a_n\omega_2\omega_1}W^{(1)}_{\omega_1\alpha_1\cdots\alpha_s}
   W^{(2)}_{\omega_2\beta_1\cdots\beta_r}.
\end{equation}
The tensor $\Delta^{a_1\cdots a_n\omega_2\omega_1}$ is
invariant by permutation of the indices $\omega_1$ and $\omega_2$ and these two factors are the same. Therefore,
when comparing $\langle W(K)\rangle$  and $\langle W(\tilde K)\rangle$ we can
use a sequence of small deformations to arrive at the configuration on figure \ref{fig10} {\bf (a)} and 
figure \ref{fig10} {\bf (b)}. The flip itself will not give any contribution and we can use the result for 
planar loops given in (\ref{eq:area-law}).

\begin{figure}[h!]
	\begin{center}
		\includegraphics[scale=1]{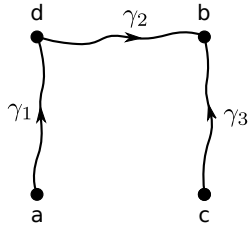}
		\caption{The paths $\gamma_1$, $\gamma_2$ and $\gamma_3$ are some paths that connect the vertices $(a,d)$, $(d,b)$ and $(b,c)$.}
		\label{fig13}
	\end{center}
\end{figure}

As for the generic case we can proceed as 
follows. Choose curves $\gamma_1,\gamma_2$  and $\gamma_3$ connecting the pairs 
$(a,d),(d,b)$ and $(c,b)$ as in figure \ref{fig13}. Since the 3-ball is simple connected, 
$\gamma_{ab}$ is isotopic to $\gamma_1\circ \gamma_2$ and $\gamma_{cd}$ can be
deformed to $\gamma_3\circ \gamma^{-1}_2$. In a similar way, we have 
$\tilde\gamma_{ab}$ is isotopic to $\gamma_1\circ \gamma_2$ and $\tilde\gamma_{cd}$ can be
deformed to $\gamma_3\circ \gamma^{-1}_2$. In a similar fashion as for the particular case of figure \ref{fig10} {\bf (a)},
there will be two sequence of flaps along $\gamma_2$. One comes from the the deformation of
$\gamma_{ab}$ and the other comes from the deformation $\gamma_{cd}$.  That has to be compared with a similar sequence of flaps
coming from the deformation of $\tilde\gamma_{ab}$ and
$\tilde\gamma_{cd}$. As in the case of figure \ref{fig12} {\bf (c)}
and {\bf (d)}, 
these sequences of pairs of flaps along $\gamma_2$ can only differ by a permutation. Since $\Delta^{a_1\cdots a_n}$ is symmetric by 
permutation of indices we can conclude that 
flips do not give any contribution and we can use the result for the 
planar loops given in (\ref{eq:area-law}) for any knot $K$. 

\section{Final Remarks}\label{sec:final}

Topological field theories are among the simplest lattice models we can have. 
From the physics point of view they are peculiar models. Partition function
and correlations can be computed but the dynamics is too simple. Rather than
considering TQFTs in isolation, we have looked at the problem from a broad perspective
and investigated a two parameter family of models where TQFTs can arise at certain
points of the parameter space. We have considered gauge theories with $\mathbb{Z}_2$
symmetry since it is the simplest gauge group but can still accommodate 
non trivial models, such as the $3D$ spin-gauge model. These more familiar models appear as one parameter curves in the two dimensional parameter space $\Gamma$.

We have found several limits that we can loosely call  topological or quasi-topological
comprising a subset $\Gamma_0$ of $\Gamma$. 
On $\Gamma_0$ both partition function and expectation value of Wilson loops were computed.
The partition function points on
$\Gamma_0$ are  topological numbers up to an overall scale factor. One could think that
$\Gamma_0$ contain only topological models but the expectation value of the Wilson reveals
something else. First of all, $\langle W_R(\gamma)\rangle$ does not depend on the isotopy class
of the curve $\gamma$. 
Furthermore, for a subset of $\Gamma_0$, 
$\langle W_R(\gamma)\rangle$ depends on the size of $\gamma$ and follows a discrete version of an
area law. 

In the parametrization $(\alpha^0,\alpha^1)$ of $\Gamma$ used in the paper, 
the subset $\Gamma_0$ is made of 4 straight lines passing through $(0,0)$.
By looking at the gauge Ising model, we can see that it approaches three of these lines
 for $\beta\rightarrow\pm \infty$ 
and $\beta \rightarrow 0 $. There is an extra line given by $\alpha^1=-\alpha^0$ that, as far as
we know, does not relate directly to any physical model. 

The existence of a set $\Gamma_0$ in the parameter space where the model behaves in
a topological way can be seen as an Euclidean version of topological order. It seems
that, rather than a special case, the same phenomena will happen for gauge theories
with any compact gauge group $G$. For $\mathbb{Z}_N$ and non abelian groups, the analysis is much more involved and it will
be reported in a separated paper.  

\section*{Acknowledgments}
The authors would like to thank A.P. Balachandran for discussions.
This work was supported by Capes, CNPq and Fapesp.

\end{document}